# Analysis of vortex breakdown in an enclosed cylinder based on the energy gradient theory

Meina Xiao[1,2,3], Hua-Shu Dou[*,1], Chuanyu Wu[1], Zuchao Zhu[1], Xifeng Zhao[2],

Songying Chen[3], Hongli Chen[1], Yikun Wei[1]

[1] *Faculty of Mechanical Engineering and Automation, Zhejiang Sci-Tech University, 310018 Hangzhou, Zhejiang, PR China*

[2] *Hisense Air-conditioner R&D Center, Hisense (Shandong) Air-conditioner Co. Ltd., 266061 Qingdao, Shandong, PR China*

[3] *Key Laboratory of High-efficiency and Clean Mechanical Manufacture, National Demonstration Center for Experimental Mechanical Engineering Education, School of Mechanical Engineering, Shandong University, 250061 Jinan, Shandong, PR China*

*Abstract*

Numerical simulation is carried out to study the phenomenon of vortex breakdown in an enclosed cylinder. The energy gradient theory is used to explain the vortex breakdown in the cylinder with consideration of centrifugal force, Coriolis force, angular momentum and azimuthal vorticity. The research results show that the large value of energy gradient function *K* is mainly located at the centerline and the region between the circulation vortices on both sides of the cylinder and the vortex breakdown bubbles at the centerline. It is found that the position of the local peak value of the energy gradient function *K* at the centerline corresponds to the location of vortex breakdown first occurrence. The position of the local peak value of *K* function in horizontal direction corresponds to the velocity inflection points except for the centerline. The vortex breakdown is mainly determined by the high *K* value at the centerline for low aspect ratio. The influence of the region of high *K* value between the circulation vortices on both sides of the cylinder and the vortex breakdown bubbles at the centerline becomes larger with the increase of the aspect ratio. The occurrence and development of the vortex breakdown bubble may be affected by the region of high *K* value between the circulation vortices on both sides of the cylinder and the vortex breakdown bubbles at the centerline for high aspect ratio.

______________________________________________________________________

## 1. Introduction

The vortex devices are widely used in chemical, biological, energy and other industries. Swirling flow is a very common and important phenomenon occurred in the vortex devices, such as centrifugal separator, bioreactor, pump, vortex flow meter, vortex mixer, agitator and so on. The flow characteristics of the swirling flow could influence the mass transfer and heat transfer, subsequently affects the chemical reaction, biochemical reactions and cell metabolism. It also plays an important role in the separation and purification of the product.

The vortex breakdown phenomena would occur in the closed cylinder driven by the bottom wall with the increase of Reynolds number. The vortex breakdown in the enclosed cylinder is easy to occur and easy to control. Therefore, it has attracted a lot of attention from researchers in recent years.

______________________________________________________________________

[*]Corresponding author. Tel.: +86 57186843661, Fax.: +86 57186843350.
  E-mail address: huashudou@yahoo.com (H.-S. Dou).

The control strategies can be applied to regulate beneficially the flow in a closed cylinder using various cross-section configurations of the closed container (Naumov et al., 2015; Yu et al., 2006), partial rotating lids (Piva et al., 2005; Yu et al., 2007; Jørgensen et al., 2010; Mununga et al., 2014), small rotating disk (Mununga et al., 2004; Tan et al., 2009), temperature gradient (Herrada & Shtern, 2003a, 2003b), addition of near-axis swirl modification (Herrada et al., 2003b; Husain et al., 2003; Jacono et al., 2008; Cabeza et al., 2010), tilted endwalls (Hourigan, 2013), and density effect (Ismadi et al., 2011; Adzlan et al., 2012).

Many parameters have been applied to characterize and explain the onset and the development of the vortex breakdown. The roles of centrifugal force (Brown and Lopez, 1988; Gelfgat et al., 1996; Yu et al., 2007), Coriolis force (Gelfgat et al., 1996), pressure gradient (Brown and Lopez, 1988; Husain et al., 2003; Yu et al., 2007), angular momentum (Brown and Lopez, 1988; Yu et al., 2007; Yu et al., 2008) and azimuthal vorticity (Gelfgat et al., 1996) were used to explain the phenomenon of vortex breakdown.

For the study of flow stability, the linear stability theory of parallel flow is mostly used. The further research shows that it is very important to consider the non-parallelism and nonlinearities, and the effect of the pressure gradient is also considered. In a variety of numerical methods, the method of parabolized stability equation (Bistrian et al. 2013; Bistrian 2014) has a particularly important role. In the past years, Dou and co-authors proposed the energy gradient theory which is used to analyze flow instability and turbulent transition. Currently, the energy gradient theory has been successfully used to study the plane Couette flow, pipe Poiseuille flow, plane Poiseuille flow (Dou, 2006; Dou, 2011), Taylor-Couette flow (Dou et al., 2008), natural convection (Dou et al., 2016) etc. This theory is also successfully used to analyze the instability in viscoelastic flows (Dou et al., 2007; Dou et al., 2008; Lu et al., 2006). It is found that the simulation results are in good agreement with the experimental ones in that the position with the large energy gradient function $K$ value will lose its stability earlier than that with low $K$ value.

In this paper, the energy gradient theory is applied to the numerical simulation results for further analysis of the flow stability in the cylinder. The position of the onset of the vortex breakdown and the distribution of energy gradient function $K$ are compared. The research shows that the position of the large value of $K$ at the centerline indicates the first place where the flow becomes unstable, then gives rise to the vortex breakdown bubble.

This paper is organized as follows: Section 2 describes the governing equations and numerical method. Section 3 investigates the calculation of the energy gradient function $K$. Section 4 analyzes the simulation results, including three conditions of low aspect ratio, medium aspect ratio and high aspect ratio. Section 5 summarizes the findings.

## 2. Governing equations and numerical method

*2.1 Geometric model and grid*

Vortex breakdown in an enclosed cylinder can be described by two parameters: Reynolds number $Re=\Omega r^2/\upsilon$ and aspect ratio $h/r$, where $\Omega$ is the angular velocity of the rotating bottom wall,

$h$ is the height of the cylinder, $r$ is radius of the cylinder, and $\upsilon$ is the kinematic viscosity.

GAMBIT software is used to generate the mesh that is illustrated in Fig. 2. The hexahedron elements are applied to divide the computational domain. The grid density near the axis of the cylinder is very large because the vortex breakdown would occur in this region. The number of the grid in the cylinder is 2180500, 3264000 and 4657500, respectively.

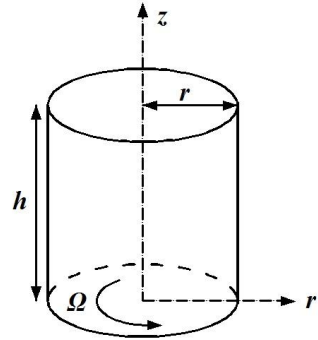

**Fig. 1.** Geometric model.

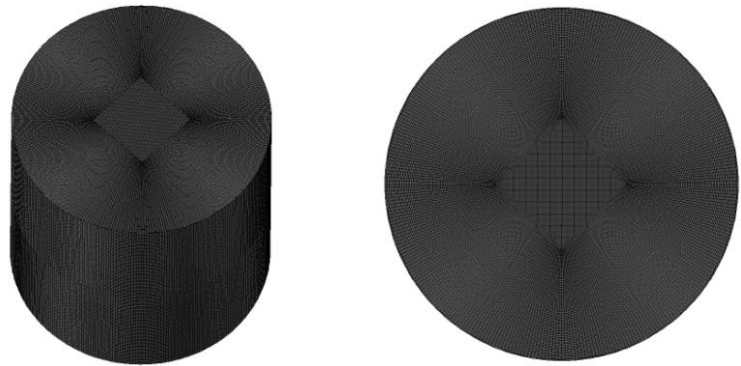

**Fig. 2.** Grid in the cylinder.

**Table 1**
Check of grid independence.

| Grid number | Axial Velocity (m/s) |
| --- | --- |
| 2.18 million | 0.010389 |
| 3.26 million | 0.010199 |
| 4.66 million | 0.010058 |

The grid independence has been verified at the condition of $Re=1431$, $H/R=2$. The variation of the axial velocity at the monitoring point $(0, 0, h/2)$ calculated with the three meshes is less than 2%, as shown in Table 1. Thus, the density of the grid has little effect on the accuracy of the calculation results when the grid number is larger than 2.18 million. To better capture the position of onset of the vortex breakdown, a grid number of 3.26 million is used. Through the grid quality examination, the grid equal angle skew and equal size skew grid is less than 0.5. So the grid quality is good.

*2.2 Governing equations*

Both steady and unsteady flow equations have been used for numerical modeling of the onset of vortex breakdown phenomena (Yu et al., 2006, 2007, 2008; Escudier et al., 2007). In present study, the Reynolds number is less than 2500, so the flow in the cylinder is steady (Escudier, 1984). The base flow is laminar due to the low Reynolds number. The governing equations are the steady Navier-Stokes equations in Cylindrical coordinates that are expressed as follows:

$$\frac{1}{r}\frac{\partial(rv_r)}{\partial r}+\frac{\partial v_z}{\partial z}=0 \tag{1}$$

$$\frac{1}{r}\frac{\partial(rv_r v_r)}{\partial r}+\frac{\partial(v_z v_r)}{\partial z}=-\frac{1}{\rho}\frac{\partial p}{\partial r}+\upsilon\left[\frac{\partial}{\partial r}\left(\frac{1}{r}\frac{\partial}{\partial r}(rv_r)\right)+\frac{\partial^2 v_r}{\partial z^2}\right]+\frac{v_\theta^2}{r} \tag{2}$$

$$\frac{1}{r}\frac{\partial(rv_r v_z)}{\partial r}+\frac{\partial(v_z v_z)}{\partial z}=-\frac{1}{\rho}\frac{\partial p}{\partial z}+\upsilon\left[\frac{1}{r}\frac{\partial}{\partial r}\left(r\frac{\partial v_z}{\partial r}\right)+\frac{\partial^2 v_z}{\partial z^2}\right] \tag{3}$$

$$\frac{1}{r}\frac{\partial(rv_r v_\theta)}{\partial r}+\frac{\partial(v_z v_\theta)}{\partial z}=\upsilon\left[\frac{\partial}{\partial r}\left(\frac{1}{r}\frac{\partial}{\partial r}(rv_\theta)\right)+\frac{\partial^2 v_\theta}{\partial z^2}\right]-\frac{v_r v_\theta}{r} \tag{4}$$

where $v_r$, $v_\theta$ and $v_z$ are the radial, azimuthal, and axial velocities, respectively; $r$ and $z$ are the radial and axial coordinates, respectively.

*2.3 Numerical method*

The governing equations are solved in this paper by the finite volume method. The convective term in the governing equations is discretized with the second order upwinding scheme, while the diffusion term is approximated with the central difference scheme. The discretized equations are solved using the Semi-Implicit Method for Pressure-Linked Equation (SIMPLE) method. The convergence criterion is set as the residual of the each variable to be less than $10^{-6}$. The boundary conditions are as follows. The bottom wall of the cylinder rotates around the $z$ axis at the angular velocity $\Omega$. No-slip and no-penetration boundary condition is employed on all the walls.

*2.4 Validation of numerical calculation*

(a) *Re*=1492                    (b) *Re*=1854

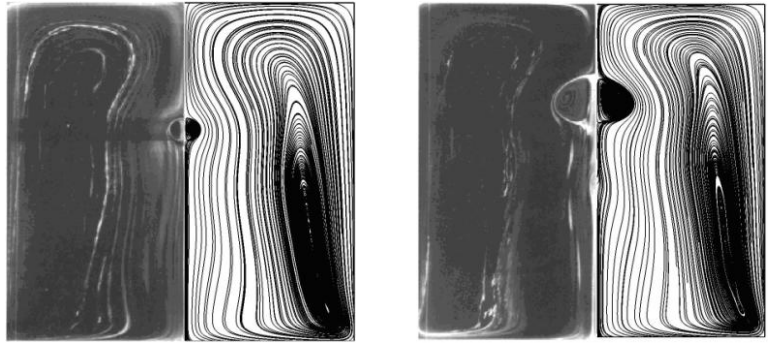

**Fig. 3.** Comparison of experimental data and numerical results of the flow structure for aspect ratio $h/r=2$ and $Re=1492$, $Re=1854$ (**Fig. 2** in reference (Escudier, 1984)).

In order to validate the employed numerical method, comparison is made between the numerical results and the experimental ones (Escudier, 1984). The flow visualization results in reference (Escudier, 1984) are shown on the left sides of Fig. 3 and the streamlines of the numerical simulation are showed on the right sides. The agreement between experimental data and numerical result shows the accuracy of the numerical calculation results.

## 3. Calculation of the energy gradient function $K$

For a given incompressible base flow perturbed by a disturbance, a stability criterion can be written as follows (The derivation of equation (5) can be found in Dou, 2011),

$$K \frac{v'_m}{v} < Const \tag{5}$$

where $K$ is a dimensionless variable and is a function of coordinates; $v$ is the velocity of main flow; $v_m$ is the amplitude of the disturbance of velocity. The dimensionless variable $K$ can be considered as a local Reynolds number. In laminar flow, its magnitude represents the extent of a flow approaching the instability and turbulent transition (Dou, 2006; Dou, 2011).

For situations where both pressure driven flow and shear driven flow are present simultaneously, a universal equation for calculating the energy gradient function is obtained.

$$K = \frac{v \frac{\partial E}{\partial n}}{v \frac{\partial W}{\partial s}} = \frac{v \frac{\partial E}{\partial n}}{v \frac{\partial E}{\partial s} + \phi} \tag{6}$$

Here, $W$ is the work done by the viscous shear stress; $E = p + \rho v^2 / 2$ is the total mechanical energy; $\rho$ is the fluid density; $s$ is the streamwise direction and $n$ is the transverse direction of the local streamline. Further, $\varphi$ is the energy dissipation function caused by the viscosity and it can be written as,

$$\phi = 2\mu D^2 = 2\mu \left[ \frac{1}{2} \left( \nabla V + \nabla V^T \right) \right]^2 \tag{7}$$

Here, *D* is the strain rate tensor; $\mu$ is the dynamic viscosity coefficient.

It can be found from equations (5) and (6) that the distribution of the *K* value determines the flow instability for given disturbance. In flow field, the instability would occur firstly at the location of the local peak value of *K*. The equation (6) is derived from Navier-Stokes equations and for the detailed derivations sees the reference (Dou, 2014, 2016).

## 4. Simulation results and discussions

### 4.1 Calculation of critical condition

Different critical flow conditions for *h/r*=1.22, 1.5, 2.0 and 2.5 are compared in Fig. 4. Here *h* and *r* are the height and radius of the cylindrical chamber, respectively. This well agreement verifies that the present numerical method produce reliable results and thus can be used to investigate the occurrence and the development of vortex breakdown in the following studies.

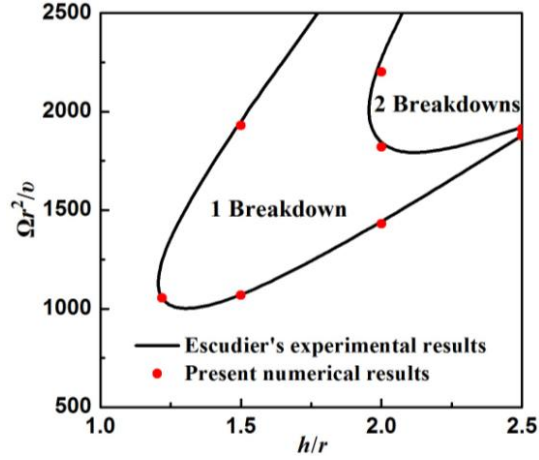

**Fig. 4.** Flow critical conditions for different breakdowns in the ($\Omega r^2/\upsilon$, *h/r*) plane (**Fig. 7** in reference (Escudier, 1984)).

### 4.2 A new interpretation for vortex breakdown

The phenomenon of vortex breakdown has been illustrated in terms of centrifugal force, Coriolis force, pressure gradient, angular momentum and azimuthal vorticity.

The relative motion equation in rotating coordinate system is written as follows:

$$\frac{\partial \vec{w}}{\partial t} + \vec{w} \cdot \nabla \vec{w} = -\frac{\nabla p}{\rho} + 2\vec{\omega} \times \vec{w} - \vec{\omega} \times (\vec{\omega} \times \vec{r}) + \frac{\mu}{\rho} \nabla^2 \vec{w} \tag{8}$$

Here, *w* is relative velocity, $\omega$ is angular velocity. The first three items on the right side of equation (8) are pressure gradient, Coriolis force and centrifugal force, respectively. Therefore, the phenomenon of vortex breakdown explained by the pressure gradient, Coriolis force and centrifugal force is based on the basic motion equation. The formation and development of vortex breakdown are discussed through the force analysis of the fluid particles.

The energy gradient function $K$ reflects influences of $(\partial E/\partial s)$ and $(\partial E/\partial n)$, not only effect of $(\partial p/\partial s)$, since the instability is not dependent on $(\partial p/\partial s)$. At the stagnation point in this study, $\partial E/\partial s = \partial p/\partial s$ is a special case. In the energy gradient theory, the negative gradient of total pressure at streamwise direction helps the flow stability, the magnitude of the gradient of total pressure at transverse direction promotes the flow stability.

The angular momentum $\Gamma$ and azimuthal vorticity $\eta$ are written as,

$$\Gamma = v_\theta r \tag{9}$$

$$\eta = \frac{\partial v_r}{\partial z} - \frac{\partial v_z}{\partial r} \tag{10}$$

Comparing the respective expressions of the energy gradient function $K$, angular momentum $\Gamma$ and azimuthal vorticity $\eta$ in equations (6), (9) and (10), it is found that the expression of the angular momentum does not consider the influence of the viscosity. The azimuthal vorticity expresses the influence of the vorticity, and the effect of viscosity is not included either. The energy gradient function considered the influences of the vorticity and the viscosity simultaneously in the expression. Therefore, the energy gradient function $K$ can better describe the flow field in the cylinder and may be able to explain the phenomenon of vortex breakdown.

*4.3 Low aspect ratio*

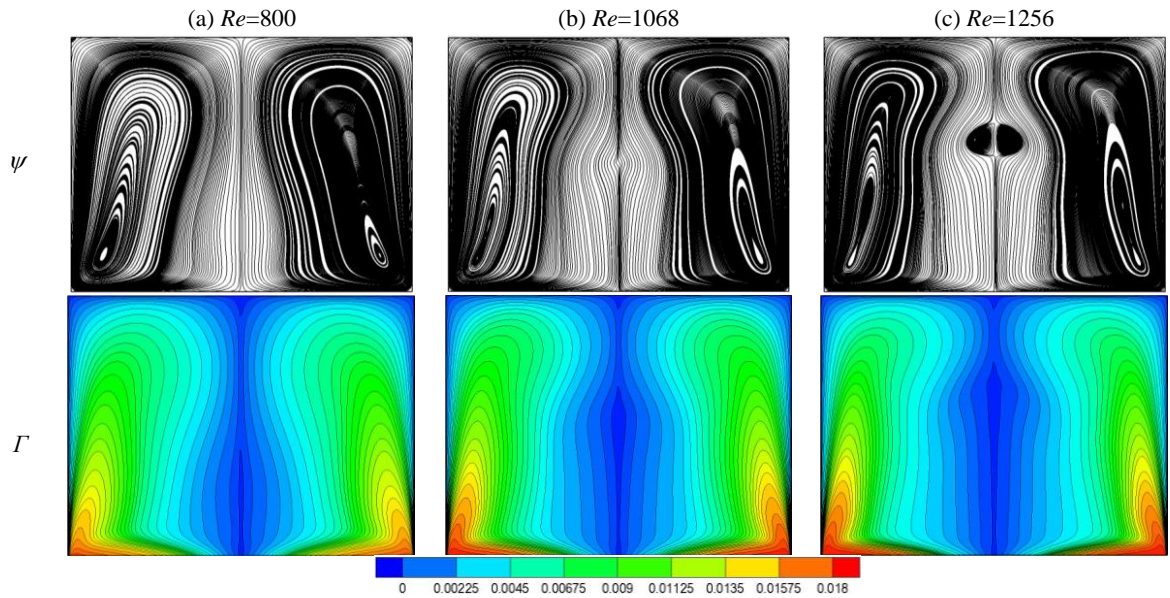

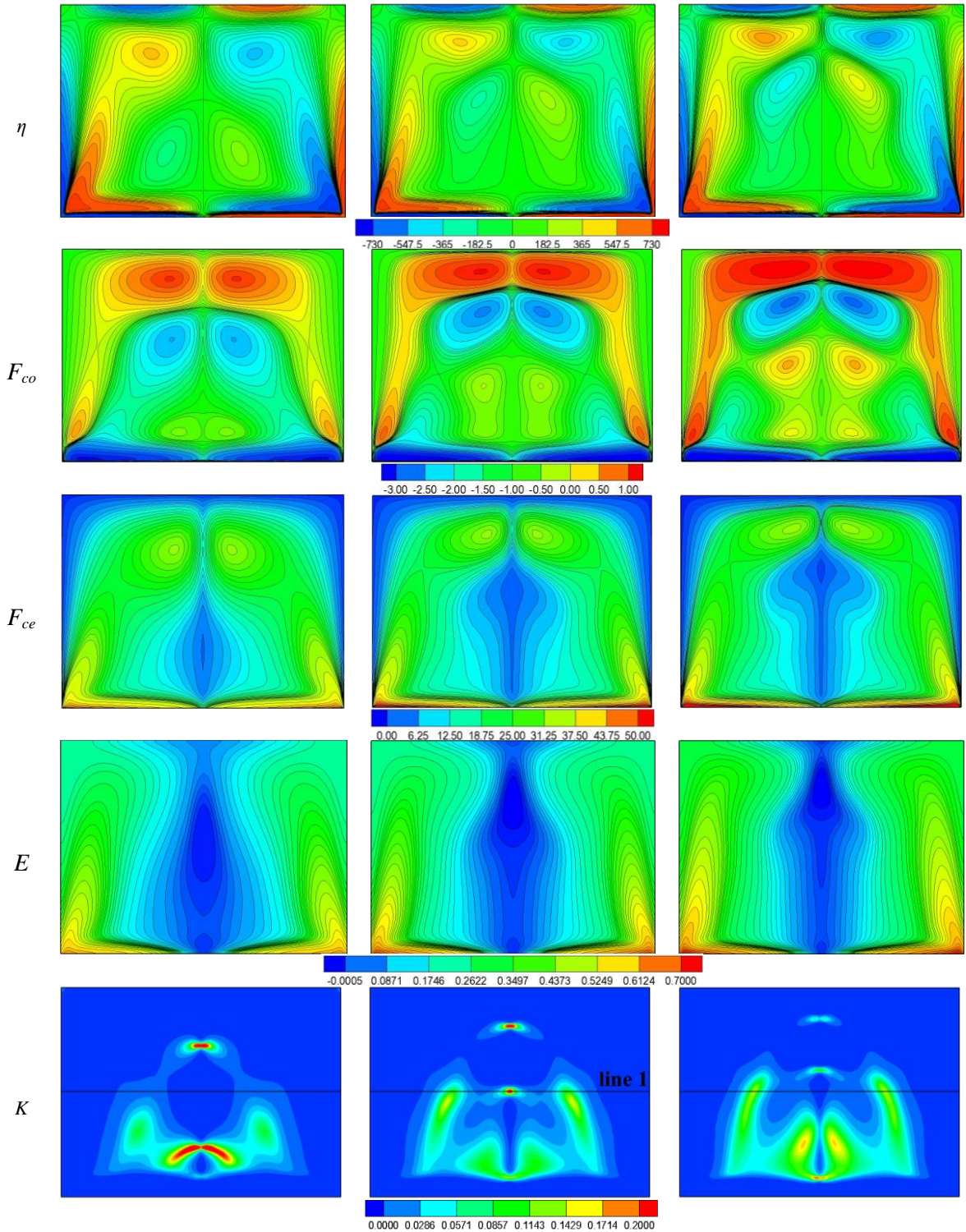

**Fig. 5.** Contours of streamlines $\psi$, angular momentum $\Gamma$, azimuthal vorticity $\eta$, Coriolis force $F_{co}$, centrifugal force $F_{ce}$, total mechanical energy $E$ and energy gradient function $K$ in the meridional plane for aspect ratio $h/r$=1.5, (a) $Re$=800; (b) $Re$=1068; (c) $Re$=1256.

In the following, the energy gradient method is employed to analyze the stability of the flow in the chamber. A detailed comparison between the contours of the streamlines $\psi$, angular momentum $\Gamma$, azimuthal vorticity $\eta$, Coriolis force $F_{co}$, centrifugal force $F_{ce}$, total mechanical energy $E$ and energy gradient function $K$ for $h/r$=1.5 in the two-dimensional meridional plane is showed in Fig. 5. The angular momentum and azimuthal vorticity both increase with the increase of $Re$. The value of the angular momentum is almost the same along a streamline (Yu et al., 2008) in Fig. 5. The position of the azimuthal vorticity near the axis moves towards the top wall. The Coriolis force acts in the rotation direction and increases the azimuthal velocity. The centrifugal force has a positive region near the top axial region, which makes the angular momentum contours and streamlines deviate away from the axis there (Yu et al., 2008; Gelfgat et al., 1996). The low total pressure zone is located near the axis. It can be seen that the streamlines near the bottom axis region slopes along the radial direction at $Re$=800 in Fig. 5(a). The region of the high $K$ value in the meridional plane at $Re$=800 is mainly located at the centerline of $z$=0.228$h$ and $z$=0.721$h$. Brown and Lopez (1988) pointed out that the wavy motion is initiated at downstream of the viscous boundary layer. This phenomenon can be explained by the high value of $K$ at the bottom axis region. The change of the streamlines is more obvious as $Re$ is increased to 1068 which is the critical Reynolds number between no vortex breakdown and one vortex breakdown in Fig. 5(b). A wave with a bulge is presented at the centerline of approximately $z$=0.5$h$. Comparing the streamlines $\psi$ and the contours of energy gradient function $K$ at $Re$=1068, the region of the local peak value of $K$ at the centerline is consistent with the position of the first occurrence of vortex breakdown. $Re$ increases to 1256, a vortex breakdown bubble is formed. The size of the vortex increases in Fig. 5(c). The distribution of the energy gradient function $K$ at $Re$=1256 is similar to that at $Re$=1068, while the value of $K$ decreases and the position of the high value of $K$ moves upward. Therefore, it is seen that the position of local peak value of $K$ at the centerline is consistent with the location of the onset of vortex breakdown. It is concluded that the energy gradient method can predict the position of the first occurrence of vortex breakdown.

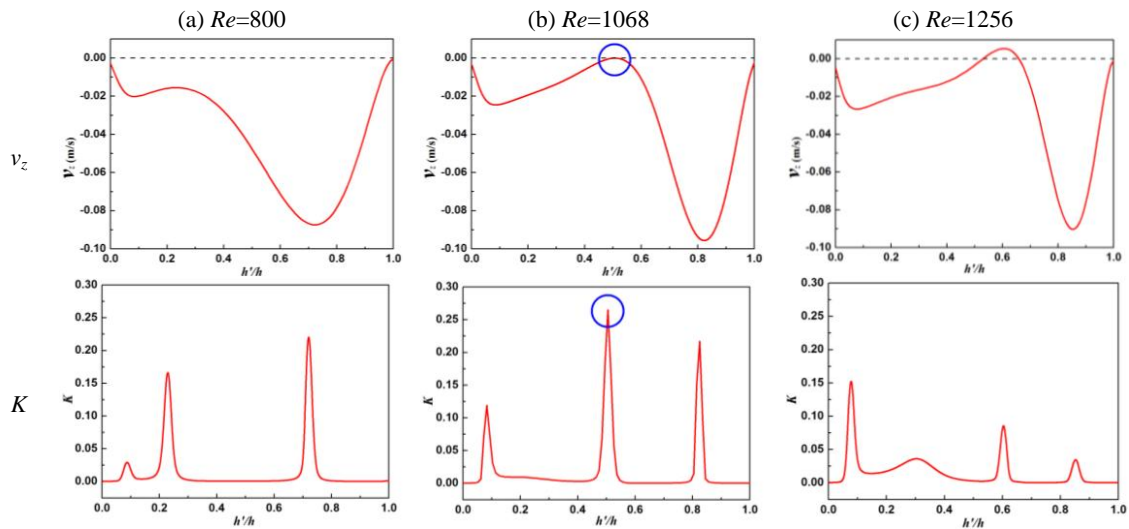

**Fig. 6.** Axial velocity and $K$ value at the centerline with the increase of $Re$ for $h/r$=1.5. (The position of local peak value of $K$ function at the centerline corresponds to the zero axial velocity gradient points.)

In order to precisely capture the first onset of the vortex breakdown bubble, it is required to compare the axial velocity at the centerline at various $Re$ because the size of the bubble is small at the initial stage of its appearance. The axial velocity and the energy gradient function $K$ at the centerline for $h/r$=1.5 are plotted in Fig. 6 with the increase of $Re$. In Fig. 6, the abscissa is the height ratio $h'/h$, and $h$ is the height of the cylindrical chamber. Comparing the axial velocity $v_z$ and the energy gradient function $K$ in Fig. 6, it is seen that the position of the local peak value of $K$ function at the centerline corresponds to the zero axial velocity gradient points. When the vortex breakdown occurs, the stagnation point appears and both the axial velocity and its gradient are zero. It can be found from Fig. 6(b) that the vortex breakdown first occurs at the centerline of $h'/h$=0.505 for $h/r$=1.5 which coincides with the position of the local peak value of the energy gradient function $K$. Here the $K$ value is larger than 0.25 in Fig. 6(b). With the further increase of $Re$ and the development of the vortex breakdown bubble, the value of $K$ at the centerline deceases rapidly.

It can be concluded from Fig. 6 that the position of the local peak value of $K$ at the centerline is consistent with the region of the onset of vortex breakdown at critical Reynolds number $Re_c$. The critical Reynolds number $Re_c$ is used to describe the flow instability and the development of vortex breakdown in the previous studies. However, the critical Reynolds number $Re_c$ is limited to show the overall flow state in the flow field and cannot interpret the exact location where the instability first occurs. In present study, the unstable zone can be indicated exactly according to the distribution of $K$ value. Therefore, the energy gradient function $K$ can better explain the onset of the vortex breakdown.

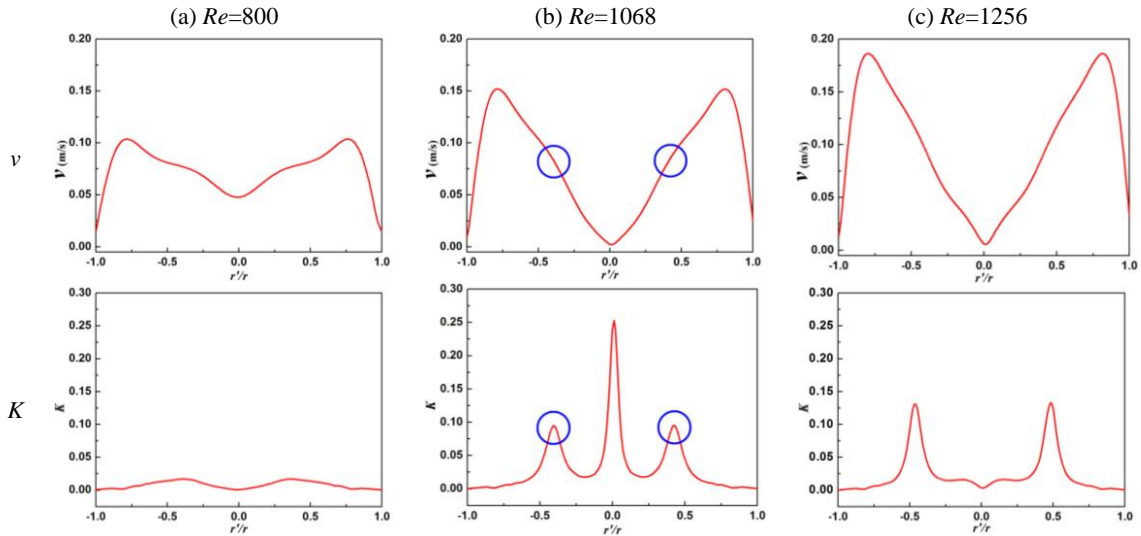

**Fig. 7.** Velocity and $K$ value along the radial direction at line 1 with the increase of $Re$ for $h/r$=1.5. (The position of local peak value of $K$ function corresponds to the velocity inflection points except at the centerline.)

In Fig. 5, a horizontal line 1 crosses the position of the first occurrence of vortex breakdown for $h/r$=1.5. The velocity and $K$ value at line 1 on the cross section of the cylinder are plotted in

Fig. 7 with the increase of *Re*. In Fig. 7, the abscissa is the radius ratio *r'/r*, and *r* is the radius of the cylindrical chamber. In Fig.7, at Re=1068, there are 3 peaks in the *K* distribution on the line 1 of Fig. 5. The two peaks at both sides of the centerline corresponds to the velocity inflection points, while the highest peak is located at the centerline whether there is occurrence of the vortex breakdown or not. The velocity at line 1 (in Fig. 5) increases with the increase of Re. The value of *K* in the region between the circulation vortices on both sides of the cylinder and the vortex breakdown bubbles at the centerline increases with the increase of Re. Then, it shows that the flow in the mentioned above region gradually becomes unstable.

For the studied problem case here, the velocity inflection points in horizontal direction may not have direct relationship to the vortex breakdown. However, the highest value of *K* indicates the position where the instability takes place first, .i.e., the vortex breakdown at the centerline. The other two peaks of *K* indicate the two areas with local instabilities.

It can be concluded that the region of the local peak value of *K* at the centerline is consistent with the zero axial velocity gradient point. The vortex breakdown first occurs at one point of zero axial velocity gradient. It is found that the region of high *K* value between the circulation vortices and the vortex breakdown bubbles has little influence on the development of the vortex breakdown bubble for the low aspect ratio. The relationship between the *K* and velocity inflection point has been explained in previous works for other various problems studied (Dou, 2006, 2011).

*4.4 Medium aspect ratio*

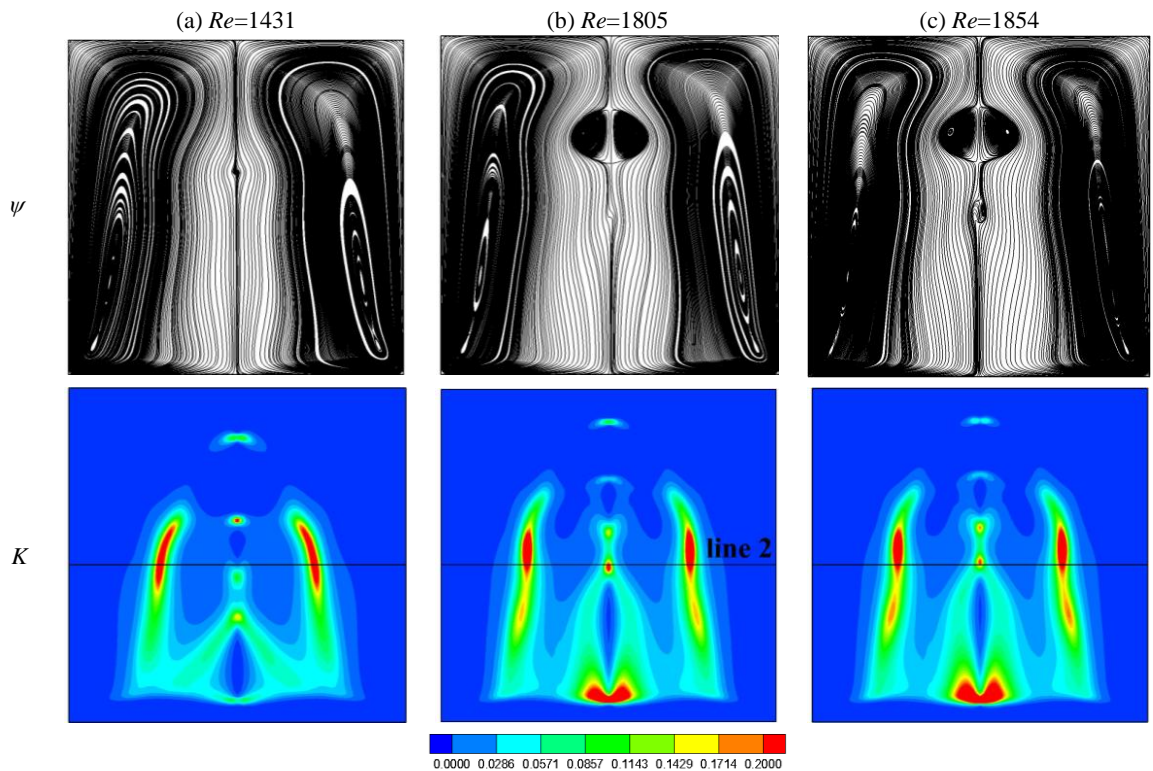

**Fig. 8.** Contours of streamlines $\psi$ and *K* in the meridional plane for *h/r*=2, (a) *Re*=1431; (b) *Re*=1805; (c) *Re*=1854.

The contours of the streamlines $\psi$ and energy gradient function $K$ for $h/r=2$ are shown in Fig. 8. From the contours of streamlines in Fig. 8, it is seen that one vortex breakdown first occurs at $Re_c=1431$ and then the second vortex breakdown appears at $Re_c=1805$ with the further increase of $Re$. The positions of the occurrence of the two vortex breakdown bubbles are consistent with the regions of the local peak value of $K$ at the centerline in Figs. 9(a) and 9(b). The $K$ value is high in the region between the circulation vortices on both sides of the cylinder and the vortex breakdown bubbles at the centerline, which shows this region has a certain influence on the development of the vortex breakdown bubble. The value of $K$ increases and the region of the high value of $K$ is extended considerably from $Re=1805$ to $Re=1854$.

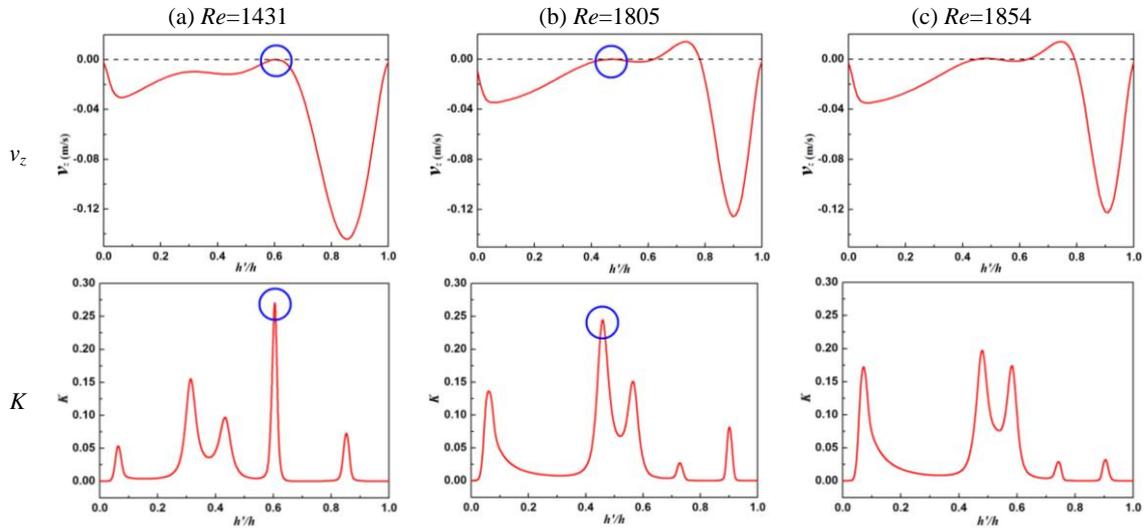

**Fig. 9.** Axial velocity and $K$ at the centerline with the increase of $Re$ for $h/r=2$.

The axial velocity and the $K$ function at the centerline for $h/r=2$ are presented in Fig. 9 with the increase of $Re$. The points of zero axial velocity suggest the onset of vortex breakdown. The critical axial velocities $v_z=0$ are located respectively at $h'/h=0.596$ for $Re_c=1431$ and at $h'/h=0.485$ for $Re_c=1805$. So the first vortex breakdown occurs at $h'/h=0.596$ and the second vortex breakdown occurs at $h'/h=0.485$. At the centerline, the position of the local peak value of $K$ function corresponds to the stagnation point. The maximum of the $K$ value in Figs. 9(a) and 9(b) both are larger than 0.25 when the vortex breakdown occurs. Then, the value of $K$ at the centerline deceases with the further increase of $Re$.

(a) $Re=1431$     (b) $Re=1805$     (c) $Re=1854$

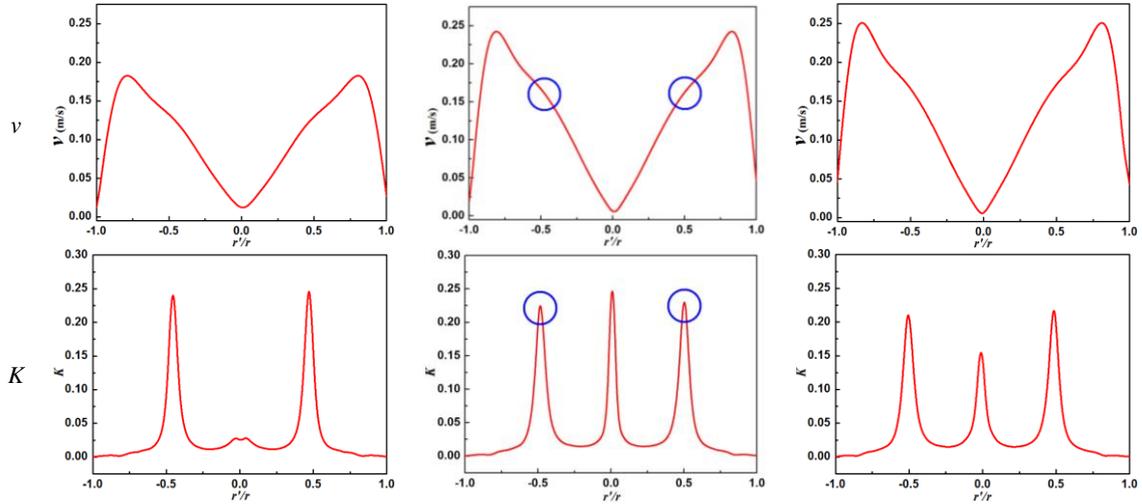

**Fig. 10.** Velocity and *K* value at line 2 with the increase of *Re* for *h/r*=2.

In Fig. 8, a horizontal line 2 in the meridional plane has the same height with the starting position of the second vortex breakdown bubble. The velocity and the *K* value at line 2 for *h/r*=2 are plotted in Fig. 10 with the increase of *Re*. It can be seen from Fig. 10 that two positions of high *K* value in the region between the circulation vortices on both sides of the cylinder and the vortex breakdown bubbles at the centerline correspond to the velocity inflection points. The value of *K* in the region between the circulation vortex and the vortex breakdown bubble is very high, which is almost as high as the local peak value of *K* at the centerline. It suggests that the influence of the above region on the development of the vortex breakdown bubble increases with the increase of the aspect ratio.

*4.5 High aspect ratio*

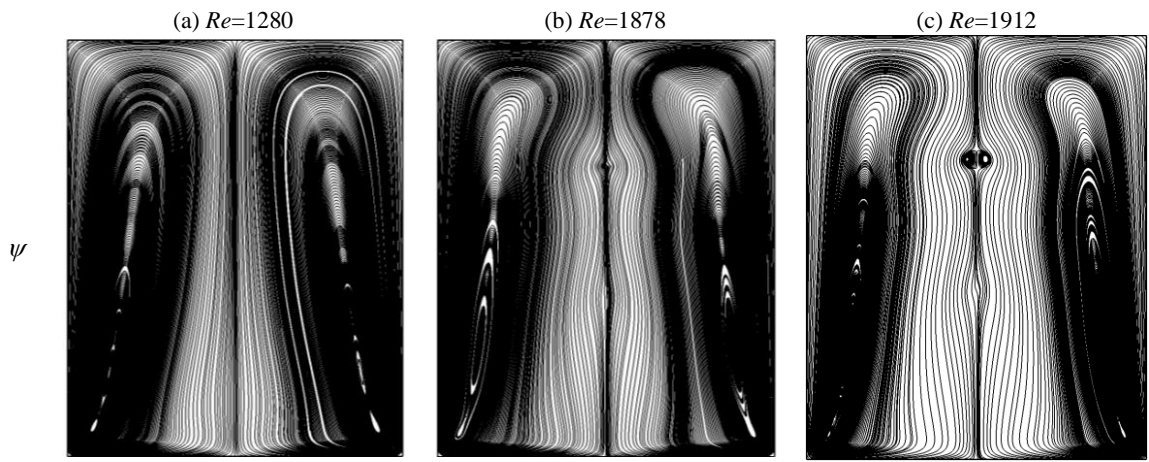

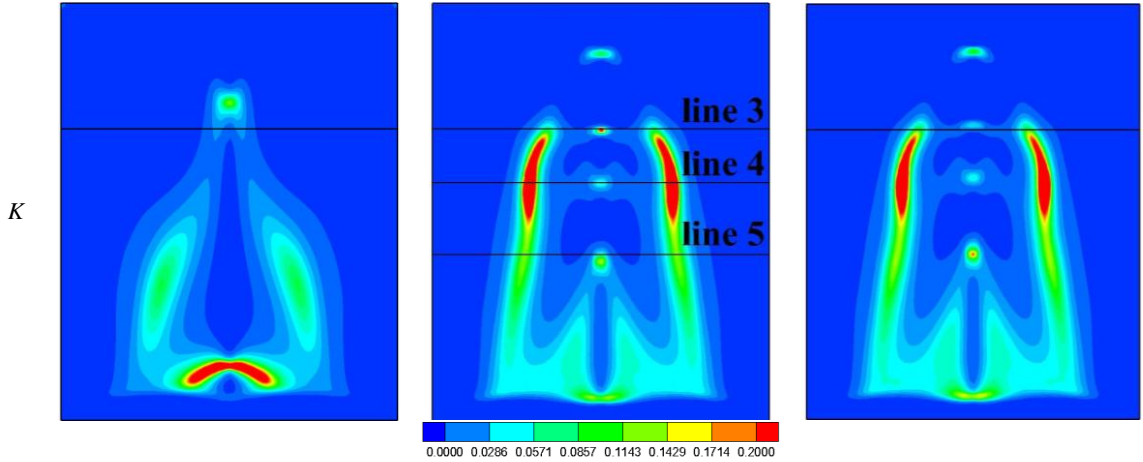

**Fig. 11.** Contours of streamlines $\psi$ and $K$ in the meridional plane for aspect ratio $h/r$=2.5, (a) $Re$=1280; (b) $Re$=1878; (c) $Re$=1912.

The contours of the streamlines $\psi$ and energy gradient function $K$ for $h/r$=2.5 are showed in Fig. 11. The development of the streamlines $\psi$ and energy gradient function $K$ for $h/r$=2.5 with the increase of $Re$ is similar to that for $h/r$=2. The first vortex breakdown and the second vortex breakdown occurs subsequently with the increase of $Re$. The positions of the beginning of the vortex breakdown are still in agreement with the points of the local peak value of $K$ at the centerline. But the local peak value of $K$ in the meridional plane isn't located at the centerline, but in the region between the circulation vortices on both sides of the cylinder and the vortex breakdown bubbles at the centerline. The region of high $K$ value in this region has a great influence on the occurrence and the development of the vortex breakdown bubble.

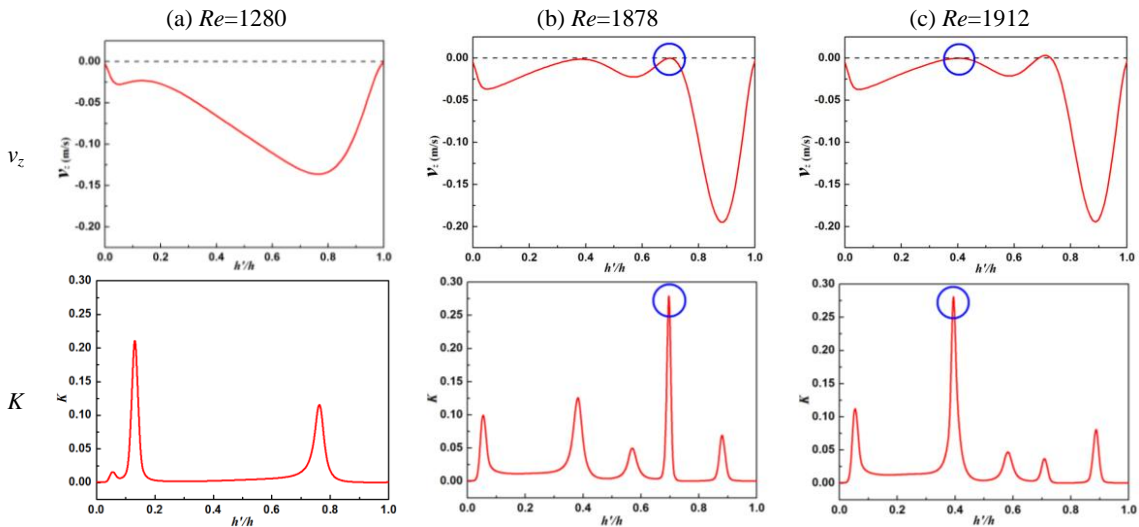

**Fig. 12.** Axial velocity and $K$ at the centerline with the increase of $Re$ for $h/r$=2.5.

Fig. 12 shows the distribution of the axial velocity and the energy gradient function $K$ at the centerline for $h/r$=2.5 with the increase of $Re$. At the centerline of the cylinder, the position of the high *K* value corresponds to the points of zero axial velocity gradient. The position of the local peak value of *K* at the centerline corresponds to region of the first occurrence of vortex breakdown in Figs. 12(b) and 12(c).

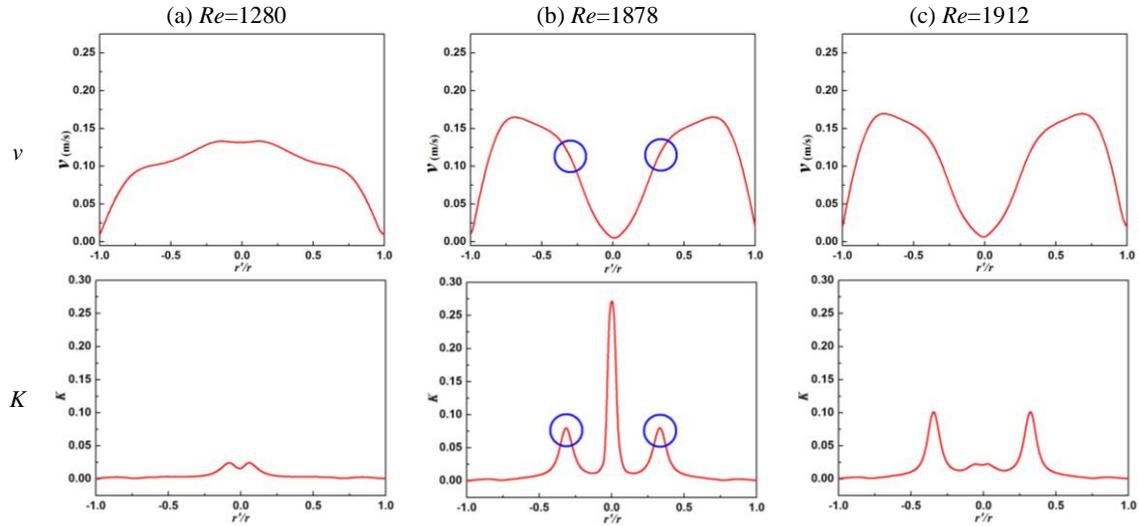

**Fig. 13.** Velocity and *K* at line 3 with the increase of *Re* for *h/r*=2.5.

In Fig. 11, the horizontal lines 3 and 5 in the meridional plane pass through the starting position where the first vortex breakdown and the second vortex breakdown occur. Line 4 lies between lines 3 and 5. The velocity and the *K* value at line 3 for *h/r*=2.5 are showed in Fig. 13 with the increase of *Re*. Fig. 14 shows the velocity and *K* at three lines at the critical *Re* of the occurrence of the first vortex breakdown. Two positions of high value of *K* in the region between the circulation vortices on both sides of the cylinder and the vortex breakdown bubbles at the centerline correspond to the velocity inflection points in Figs. 13 and 14. Here the value of *K* is very high, which is higher than the local peak value of *K* at the centerline in Fig. 14(b). The unstable region between the circulation vortices on both sides of the cylinder and the vortex breakdown bubbles at the centerline may promote or induce the production and development of the vortex breakdown bubble.

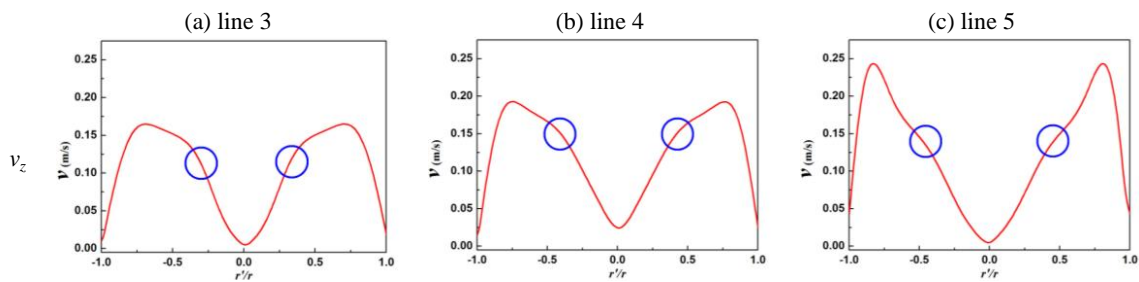

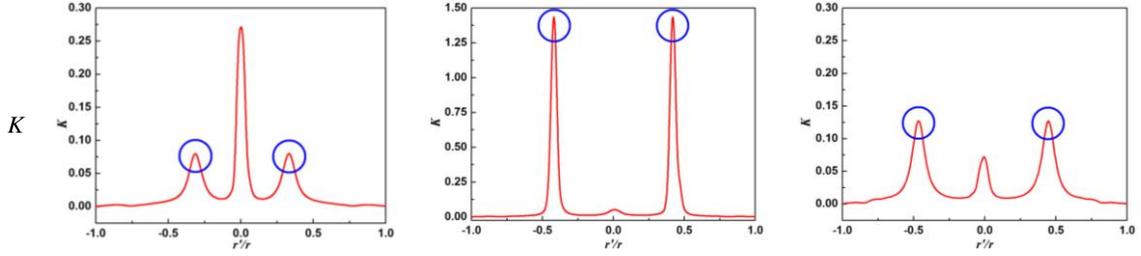

**Fig. 14.** Velocity and *K* at the horizontal lines at Re$_c$=1878 with the decrease of height for *h/r*=2.5.

From the distribution of *K* value at the centerline under different aspect ratio, it is concluded that when the local peak value of *K* is larger than 0.25, the flow near the centerline is unstable and the vortex breakdown is easy to occur.

## 5. Conclusions

Vortex breakdown of swirl flow in an enclosed cylinder with a rotating bottom wall is studied by numerical simulation. Based on the laminar base flow assumption, the incompressible, steady Navier–Stokes equations are solved for the flow in the cylinder. The function *K* calculated by the energy gradient theory is used to explain the phenomenon of vortex breakdown, compared with centrifugal force, Coriolis force, angular momentum and azimuthal vorticity. The main findings are as follows.

1. The high value of *K* is mainly located at the centerline and the region between the circulation vortices on both sides of the cylinder and the vortex breakdown bubbles at the centerline.

2. The position of the local peak value of the energy gradient function *K* at the centerline is the first place where the flow loses its stability, then causes vortex breakdown bubble. The position of local peak value of *K* function in horizontal direction corresponds to the velocity inflection points except at the centerline.

3. The influence of the region between the circulation vortices on both sides of the cylinder and the vortex breakdown bubbles at the centerline becomes larger with the increase of aspect ratio. The vortex breakdown is mainly determined by the high *K* value at the centerline for low aspect ratio, while it may be affected by region of high *K* value between the circulation vortices on both sides of the cylinder and the vortex breakdown bubbles at the centerline for high aspect ratio.

4. The distribution characteristic of *K* calculated by numerical simulation can predict the location of the vortex breakdown first occurrence. From the distribution of *K* value at the centerline under different aspect ratio, it is concluded that when the local peak value of *K* is larger than 0.25, the flow near the centerline is unstable and the vortex breakdown is easy to occur.

## Acknowledgements

This work is supported by National Natural Science Foundation of China (51536008,

51579224), Zhejiang Province Science and Technology Plan Project (2017C34007), and Zhejiang Province Key Research and Development Plan Project (2018C03046).REFERENCES

Adzlan, A., Gotoda, H., 2012. Experimental investigation of vortex breakdown in a coaxial swirling jet with a density difference. Chem. Eng. Sci. 80, 174-181.

Bistrian, D.A., 2013. Parabolized Navier–Stokes model for study the interaction between roughness structures and concentrated vortices. Phys. Fluids 25, 104103.

Bistrian, D.A., 2014. A solution of the parabolized Navier–Stokes stability model in discrete space by two-directional differential quadrature and application to swirl intense flows. Comput. Math. Appl. 68, 197-211.

Brown, L., Lopez, J. M, 1988. Axisymmetric vortex breakdown, Part 2: Physical mechanisms. J. Fluid Mech. 221, 553-576.

Cabeza, C., Sarasua, G., Marti, A.C., Bove, I., Varela, S., Usera, G., Vernet, A., 2010. Influence of coaxial cylinders on the vortex breakdown in a closed flow. Eur. J. Mech. (B/Fluids) 29 (3), 201-207.

Dou, H.-S., 2006. Mechanism of flow instability and transition to turbulence. Int. J. Nonlinear Mech. 41, 512-517.

Dou, H.-S., 2011. Physics of flow instability and turbulent transition in shear flows. Int. J. Phys. Sci. 6, 1411-1425.

Dou, H.-S., 2014. Secret hidden in Navier-Stokes equations: singularity and criterion of turbulent transition. The APS 67th Annual Meeting of the Division of Fluid Dynamics, Nov. 23-25, San Francisco, USA.

Dou, H.-S., 2016. A universal equation for calculating the energy gradient function in the energy gradient theory. https://arxiv.org/abs/1610.01517.

Dou, H.-S., Jiang, G., 2016. Numerical simulation of flow instability and heat transfer of natural convection in a differentially heated cavity. Int. J. Heat Mass Tran. 103, 370-381.

Dou, H.-S., Khoo, B.C., Yeo K.S., 2008. Instability of Taylor-Couette flow between concentric rotating cylinders. Int. J. Therm. Sci. 47, 1422-1435.

Dou, H.-S., Nhan, P.-T., 2007. Viscoelastic flows around a confined cylinder: instability and velocity inflection. Chem. Eng. Sci. 62, 3909-3929.

Dou, H.-S., Nhan, P.-T., 2008. An instability criterion for viscoelastic flow past a confined cylinder. Korea-Aust. Rheol. J. 20, 15-26.

Escudier, M.P., 1984. Observations of the flow produced in a cylindrical container by a rotating endwall. Exp. Fluids 2, 189-196.

Escudier, M.P., Oleary, J., Pooler, J., 2007. Flow produced in a conical container by a rotating endwall. Int. J. Heat Fluid Fl. 28, 1418-1428.

Gelfgat, A.Y., Bar-yoseph P.Z., Solan, A., 1996. Stability of confined swirling flow with and without vortex breakdown. J. Fluid Mech. 311, 1-36.

Herrada, M.A., Shtern, V., 2003a. Control of vortex breakdown by temperature gradients. Phys. Fluids 15, 3468-3477.

Herrada, M.A., Shtern, V., 2003b. Vortex breakdown control by adding near-axis swirl and temperature

gradients. Phys. Rev. E 68, 2021-2028

Hourigan, K., 2013. Mixing in a vortex breakdown flow. J. Fluid Mech. 731, 195-222.

Husain, H., Shtern, V., Hussain, F., 2003. Control of vortex breakdown by addition of near-axis swirl. Phys. Fluids 15, 271-279.

Ismadi, M.-Z.P., Meunier, P., Fouras, A., Hourigan, K., 2011. Experimental control of vortex breakdown by density effects. Phys. Fluids 23, 034104.

Jacono, D.L., Sørensen, J.N., Thompson, M.C., Hourigan, K., 2008. Control of vortex breakdown in a closed cylinder with a small rotating rod. J. Fluids Struct. 24, 1278-1283.

Jørgensen, B.H., Sørensen, J.N., Aubry, N., 2010. Control of vortex breakdown in a closed cylinder with a rotating lid. Theor. Comput. Fluid Dyn. 24, 483-496.

Lopez, J.M., 1990. Axisymmetric vortex breakdown, Part 1: Confined swirling flow. J. Fluid Mech. 221, 533-552.

Lu, Z., Khoo, B.C., Dou, H.-S., Nhan, P.-T., Yeo, K.S., 2006. Numerical simulation of fibre suspension flow through an axisymmetric contraction and expansion passages by Brownian configuration field method, Chem. Eng. Sci. 61, 4998-5009.

Mununga, L., Hourigan, K., Thompson, M.C., Leweke, T., 2004. Confined flow vortex breakdown control using a small disk. Phys. Fluids 16, 4750-4753.

Mununga, L., Jacono, D.L., Sørensen, J.N., Leweke, T., Thompson, M.C., Hourigan, K., 2014. Control of confined vortex breakdown with partial rotating lids. J. Fluid Mech. 738, 5-33.

Naumov, I.V., Dvoynishnikov, S.V., Kabardin, I.K., Tsoy, M.A., 2015. Vortex breakdown in closed containers with polygonal cross sections. Phys. Fluids 27, 124103.

Ni, M., Abdou, M. A., 2010. A bridge between projection methods and SIMPLE type methods for incompressible Navier–Stokes equations. Int. J. Numer. Meth. Eng. 72, 1490-1512.

Piva, M., Meiburg, E., 2005. Steady axisymmetric flow in an open cylindrical with a partially rotating bottom wall. Phys. Fluids 17, 063603.

Tan, B.T., Liow, K.Y.S., Mununga, L., Thompson, M.C., Hourigan, K., 2009. Simulation of the control of vortex breakdown in a closed cylinder using a small rotating disk. Phys. Fluids 21, 024104.

Yu, P., Lee, T.S., Zeng, Y., Low, H.T., 2006. Effects of conical lids on vortex breakdown in an enclosed cylindrical chamber. Phys. Fluids 18, 117101.

Yu, P., Lee, T.S., Zeng, Y., Low, H.T., 2007. Characterization of flow behavior in an enclosed cylinder with a partially rotating end-wall. Phys. Fluids 19, 057104.

Yu, P., Lee, T.S., Zeng, Y., Low, H.T., 2008. Steady axisymmetric flow in an enclosed conical frustum chamber with a rotating bottom wall. Phys. Fluids 20, 087103.